\documentclass[aps,prd,showpacs,floatfix,nofootinbib,superscriptaddress,twocolumn,10pt]{revtex4}
\usepackage{graphicx}% Include figure files
\usepackage{bm}% bold math
\usepackage{subfigure}
\usepackage{amssymb}
\usepackage{color,soul}
\usepackage{graphicx}
\usepackage{multirow}
\graphicspath{{figure/}}
\DeclareGraphicsExtensions{.pdf,.png,.jpg}
\setulcolor{red}

\newcommand{\beq}{\begin{equation}}
\newcommand{\eeq}{\end{equation}}
\newcommand{\beqa}{\begin{eqnarray}}
\newcommand{\eeqa}{\end{eqnarray}}

\begin{document}

\title{Constraining $U(1)_{L_{\mu}-L_{\tau}}$ charged dark matter model for muon $g-2$ anomaly with AMS-02 electron and positron data}

\author{Lei Zu}
\affiliation{Key Laboratory of Dark Matter and Space Astronomy, Purple Mountain Observatory, Chinese Academy of Sciences, Nanjing 210023, China}
\affiliation{School of Astronomy and Space Science, University of Science and Technology of China, Hefei, Anhui 230026, China}

\author{Xu Pan}
\affiliation{Key Laboratory of Dark Matter and Space Astronomy, Purple Mountain Observatory, Chinese Academy of Sciences, Nanjing 210023, China}
\affiliation{School of Astronomy and Space Science, University of Science and Technology of China, Hefei, Anhui 230026, China}

\author{Lei Feng\footnote{Corresponding author: fenglei@pmo.ac.cn}}

\affiliation{Key Laboratory of Dark Matter and Space Astronomy, Purple Mountain Observatory, Chinese Academy of Sciences, Nanjing 210023, China}
\affiliation{Joint Center for Particle, Nuclear Physics and Cosmology,  Nanjing University -- Purple Mountain Observatory,  Nanjing  210093, China}

\author{Qiang Yuan}
\affiliation{Key Laboratory of Dark Matter and Space Astronomy, Purple Mountain Observatory, Chinese Academy of Sciences, Nanjing 210023, China}
\affiliation{School of Astronomy and Space Science, University of Science and Technology of China, Hefei, Anhui 230026, China}

\author{Yi-Zhong Fan\footnote{Corresponding author: yzfan@pmo.ac.cn}}
\affiliation{Key Laboratory of Dark Matter and Space Astronomy, Purple Mountain Observatory, Chinese Academy of Sciences, Nanjing 210023, China}
\affiliation{School of Astronomy and Space Science, University of Science and Technology of China, Hefei, Anhui 230026, China}

\begin{abstract}
Very recently, the Fermi-Lab reported the new experimental combined results on the magnetic momentum of muon with a 4.2$\sigma$ discrepancy compared with the expectation of the Standard Model \cite{Fermi_Lab}. A new light gauge boson $X$ in the $L_{\mu}-L_{\tau}$ model provides a good explanation for the $g-2$ anomaly. A Dirac fermion dark matter with a large $L_{\mu}-L_{\tau}$ charge can explain both the $g-2$ anomaly and the dark matter relic density \cite{Asai_2021}. In this work, we focus on the case that the mass of the dark matter is larger than the mass of muon (i.e. $m_{\Psi} > m_{\mu}$) for which the channel $\Psi \Psi \rightarrow \mu^- \mu^+$ opens. Although the cross section $(\sigma v)_{\mu^{-}\mu^{+}}$ is smaller by a factor of $1/q_{\Psi}^2$ ($q_{\Psi}$ represents the $L_{\mu}-L_{\tau}$ charge of the dark matter) compared with the channel $\Psi\Psi \rightarrow XX  \rightarrow \nu\nu\bar{\nu}\bar{\nu}$, the resulting secondary electrons and positrons could imprint on their spectra above GeV energies due to the reacceleration effect of cosmic ray propagation. We use the AMS-02 measurements of electrons and positrons to constrain the annihilation cross section of the channel $\Psi\Psi \rightarrow \mu^{-}\mu^{+}$, which rules out part of the parameter space of the large $L_{\mu}-L_{\tau}$ charged dark matter model to account for the muon $g-2$ anomaly.
\end{abstract}


\maketitle

\section{Introduction}

Over the past decades, the Standard Model (SM) of particle physics achieved a great success. However, it still faces severe challenges when confronting some experimental anomalies, such as the dark matter (DM), the mass of neutrinos, and the muon $g-2$ anomaly\cite{Raidal_2001,Ma_2006,Lindner_2017}. Recently, the Fermi-lab reported a combined 4.2$\sigma$ discrepancy of the muon $g-2$ measurement from the SM prediction \cite{Fermi_Lab}. Such a tension raises a big challenge to the SM. Many models containing new interactions with the muon sector were proposed to explain this tension \cite{Lindner_2017,Cho_2011,Miller_2012,Stckinger_2013}. The $U(1)_{L_{\mu}-L_{\tau}}$ model, which assumes a local gauge symmetry, provides naturally a gauge boson $X$ to interact with the muon and can account for the $g-2$ anomaly \cite{Foot_1991,He_1991,Heeck_2011,Bi_2009}. This well-studied $L_{\mu}-L_{\tau}$ model with an MeV scale boson could also avoid the constraints from other experiments \cite{BaBar_2016,CCFR_1991,Kaneta_2017}. 
  
%Dark matter(DM) is also an unresolved mystery in modern physics, which challenges the SM since there is no DM candidates in this theory. Although the interaction between DM and the baryons is expected very weak, the interaction within the dark sector has not been ruled out\cite{Zu_mirror_2021,Zu_plasma_2021,Biswas_2016}. Thus it seems natural to extend the $L_{\mu}-L_{\tau}$ model with a charged DM under the $U(1)_{L_{\mu}-L_{\tau}}$  group. This model has been widely studied to explain the DM relic density, the mass of neutrinos, the Galactic centre gamma ray excess, and the muon g-2 anormaly \cite{Biswas_2016,Patra_2017,Han_2019}. 
  
The $L_{\mu}-L_{\tau}$ model has also been widely studied to explain the relic density of DM, the mass of neutrinos, and the Galactic centre gamma ray excess \cite{Biswas_2016,Patra_2017,Han_2019}. However, usually a heavy mass of $X$ is needed which was inconsistent with other experiments \cite{CCFR_1991, BaBar_2016, Kaneta_2017} . Recently, Ref.~\cite{Asai_2021} proposed an SM singlet Dirac DM ($\Psi$) model with an $L_{\mu}-L_{\tau}$ charge $q_{\Psi}$. Distinct from the large mass of $O(10^3)$ GeV of $X$ and DM requied to explain the relic density in \cite{Patra_2017,Biswas_2016}, this additional $q_{\Psi}$ parameter can explain the DM relic density with much lighter $X$ and DM particles. The cross section for $\Psi\Psi \rightarrow \mu^{-}\mu^{+}$ is smaller than $\Psi\Psi \rightarrow XX \rightarrow \nu\nu \bar{\nu}\bar{\nu}$ by a factor $q_{\Psi}^2$ for $m_{\Psi} > m_{X}$. The free parameter $q_{\Psi}$ opens a new window to explain both the g-2 anomaly and the DM relic density, which is also consistent with other experiment limits. Following \cite{Asai_2021}, we focus on the case $m_{\Psi}>m_{\mu}>m_{X}$, in which the channel $\Psi \Psi \rightarrow \mu^{-}\mu^{+}$ opens for the non-relativistic DM. Although the mass of DM can be small (e.g., $<$GeV), we argue that the secondary positrons and electrons from DM annihilation could also have a non-negligible effect on the GeV cosmic electron and positron spectrum due to the reacceleration effect, which can thus be probed by space-based detector like AMS-02. Parts of the parameter space to explain both the DM relic density and $g-2$ anomaly would be ruled out when considering the limits from AMS-02 data.

This work is organized as follows: In Section II, we introduce the $U(1)_{L_{\mu}-L_{\tau}}$ model that we used. In Section III we briefly describe the propagation and background of electrons and positrons. In Section IV, we set our constraints using the AMS-02 data. We conclude this work in Section V.

\section{$L_{\mu}-L_{\tau}$ Model}

In this work, we have considered an extension of the SM with a simple extra local $U(1)_{L_{\mu}-L_{\tau}}$ symmetry to the SM Lagrangian. 
Therefore, the Lagrangian remains invariant under the $SU(3)_c \times SU(2)_{L} \times U(1) \times U(1)_{L_{\mu}-L_{\tau}}$ gauge symmetry. 
The DM considered here contains an additional charge $q_{\Psi}$. The charge $q_{\Psi}$ for muon (tau) is +1(-1) \cite{Asai_2021}. Although this large charge $q_\Psi$ seems unnatural for the theory, it is allowed phenomenologically. The Lagrangian is:

\begin{eqnarray}
{\mathcal L}={\mathcal L}_{\rm SM}&-&g_{X}X_{\lambda}(\bar{\mu}\gamma^{\lambda}\mu-\bar{\tau}\gamma^{\lambda}\tau+\bar{\nu_{\mu L}}\gamma^{\lambda}\nu_{\mu L}-\bar{\nu_{\tau L}}\gamma^{\lambda}\nu_{\tau L}) \nonumber\\
&-&\frac{1}{4}X_{\mu\nu}X^{\mu\nu}+\frac{1}{2}m_X^{2}X_{\mu}X^{\mu} \nonumber\\
&+&\bar{\Psi}(i\not{\partial}-m_{\Psi})\Psi-q_{\Psi}g_XX_{\lambda}\bar{\Psi}\gamma^{\lambda}\Psi ,
\label{Lagrangian}
\end{eqnarray}
where $X^{\mu}$ and $\Psi$ denotes the $U(1)_{L_{\mu}-L_{\tau}}$ gauge boson and Dirac DM, $X^{\mu \nu}$ is the field strength of $X^{\mu}$. We have ignored the kinetic mixing and the right-hand neutrino terms since they are irrelevant for our phenomenological discussion below. Therefore we have four free parameters in this model: $m_X$, $g_X$, $m_{\Psi}$, $q_{\Psi}$. 

The new combined result on the magnetic moment of muon measured by the Fermi-Lab shows a 4.2 $\sigma$ deviation from the SM
\begin{equation}
\Delta a_{\mu}=a_{\mu}^{exp}-a_{\mu}^{SM}=251 \pm 59 \times 10^{-11}
\end{equation}
The new gauge boson $X$ could contribute to an extra magnetic moment of muon $a_\mu$. The one-loop contribution is
\begin{equation}
\Delta a_{\mu}^X=\frac{g_X^2}{8\pi^2} \int_0^1 dx \frac{2m_{\mu}^2 x^2(1-x)}{x^2m_{\mu}^2+(1-x)m_{X}^2}.
\end{equation}
As discussed in \cite{Asai_2021,Lindner_2017}, with $g_X \sim 10^{-4}$ and $m_{X} \sim $ O(10) MeV, this gauge boson $X$ could explain the $g-2$ anomaly and avoid the current experimental limits (see Fig.~1 in \cite{Asai_2021} and Fig.~32 in \cite{Lindner_2017}).

In this work, we focus on the case of $m_{\Psi}>m_{\mu}>m_X$. Thus DM could annihilate through $\Psi \Psi \rightarrow XX $ process following with $X \rightarrow \nu \bar{\nu}$. The cross section is
\begin{equation}
(\sigma v)_{XX}=\frac{(q_{\Psi}g_{X})^4}{4\pi m_{\Psi}}\frac{(m_{\Psi}^2-m_{X}^2)^{3/2}}{(2m_{\Psi}^2-m_{X}^2)^2}.
\label{sv_xx}
\end{equation}
This cross section is related to the s-channel process $\Psi \Psi \rightarrow \mu \mu$ (for $m_X \ll m_{\Psi}$) as
\begin{equation}
(\sigma v)_{\mu^{-}\mu^{+}}= \frac{1+m_{\mu}^2/(2m_{\Psi}^2)](1-m_{\mu}^2/m_{\Psi}^2)^{1/2}}{q_{\Psi}^2}\times (\sigma v)_{XX}.
\label{sv_ff}
\end{equation}
Roughly $(\sigma v)_{\mu^{-}\mu^{+}}$ is $q_{\Psi}^{-2}$ smaller than $(\sigma v)_{XX}$. The additional parameter $q_\Psi$ is helpful to explain the DM abundance and can avoid the constraints from the cosmic microwave background (CMB) observations, $(\sigma v)_{f\bar{f}}/(2m_\Psi) \leq 5.1 \times 10^{-27}~{\rm cm}^{3}~{\rm s}^{-1}~{\rm GeV}^{-1}$ \cite{Slatyer_2016,Leane_2018}\footnote{Following \cite{Asai_2021,Slatyer_2016,Leane_2018}, here we adopt the conservative limit on the cross section with charged final states.}. The typical cross section to explain the DM abundance is \cite{Saikawa_2020}
\begin{equation}
(\sigma v)/2 \simeq 3 \times 10^{-26}~{\rm cm}^3~{\rm s}^{-1}.
\end{equation}
Therefore $q_{\Psi}$ needs to be large enough that the dominating channel $\Psi \Psi \rightarrow XX \rightarrow \nu\bar{\nu}$ could reach the value required to give the correct DM abundance and the $\Psi \Psi \rightarrow f\bar{f}$ channel is consistent with the CMB limits.

In the case $m_{\Psi} > m_{\mu}$, the non-relativistic $\Psi$ would also annihilate into $\mu \bar{\mu}$ in the Milky Way and contribute to the spectrum of cosmic ray electrons and positrons. After reacceleration in the propagation, the sub-GeV $e^{+}e^{-}$ could be accelerated to higher energies \cite{Boudaud_2017}, which are detectable for the space-based experiments like AMS-02. 
%In this work, we, for the first time, calculate the impact of this DM secondary $e^{+}/e^{-}$ source in the AMS-02 data and rule out part of the favored region to explain both the DM abundance and g-2 anormaly.

\section{Cosmic ray electrons and positrons}
\subsection{Propagation}
%DM in the Milky way annihilates to $\mu \bar{\mu}$ through s channel then decays to electron and positron, which could be detected by AMS-02 after propagation. 
Cosmic ray electrons and positrons propagate diffusively in the Galaxy. Numerical tools have been developed to calculate the propagation of cosmic rays, such as GALPROP \cite{Strong_1998} and DRAGON \cite{Evoli_2008}. In this work we adopt the LikeDM code \cite{Huang_2017} to calculate the propagation process. This package employs a Green's function method based on numerical tables obtained with GALPROP for given distribution of the source. This method has been verified to be a good approximation to the GALPROP result, and is much more efficient. The propagation framework is assumed to be diffusion plus reacceleration, which was found to be well consistent with the secondary and primary nuclei measured by AMS-02 \cite{Yuan_2017,Yuan_2020}. The propagation parameters we used include the diffusion coefficient $D(E)=\beta D_{0}(E/4~{\rm GeV})^{\delta}$ with $D_{0}=1 \times 10^{29}~{\rm cm}^{2}~{\rm s}^{-1}$ and $\delta =0.33$, the Alfvenic speed which characterizes the reacceleration effect $v_{A}=26.3$~km~s$^{-1}$ \cite{Ackermann_2012}. 
%This set of propagation parameters is widely adopted as the canonical medium one. 
Low-energy cosmic rays are affected by the solar modulation. We adopt the simple force-field approximation with the modulation potential to calculate this effect \cite{Gleeson_1968}. The modulation potential we adopt is 0.6~GV.
For the DM density profile, we adopt the typical NFW distribution \cite{Navarro_1997} with local density $\rho_{0} = 0.3~GeV~{\rm cm}^{-3}$  \cite{Bertone_2009}. The injected $e^+e^-$ spectrum from $\Psi\Psi \rightarrow \mu^{-}\mu^{+}$ is calculated using the PPPC4 package \cite{Cirelli_2011}.

\begin{figure}[htbp]
\includegraphics[width=1\columnwidth]{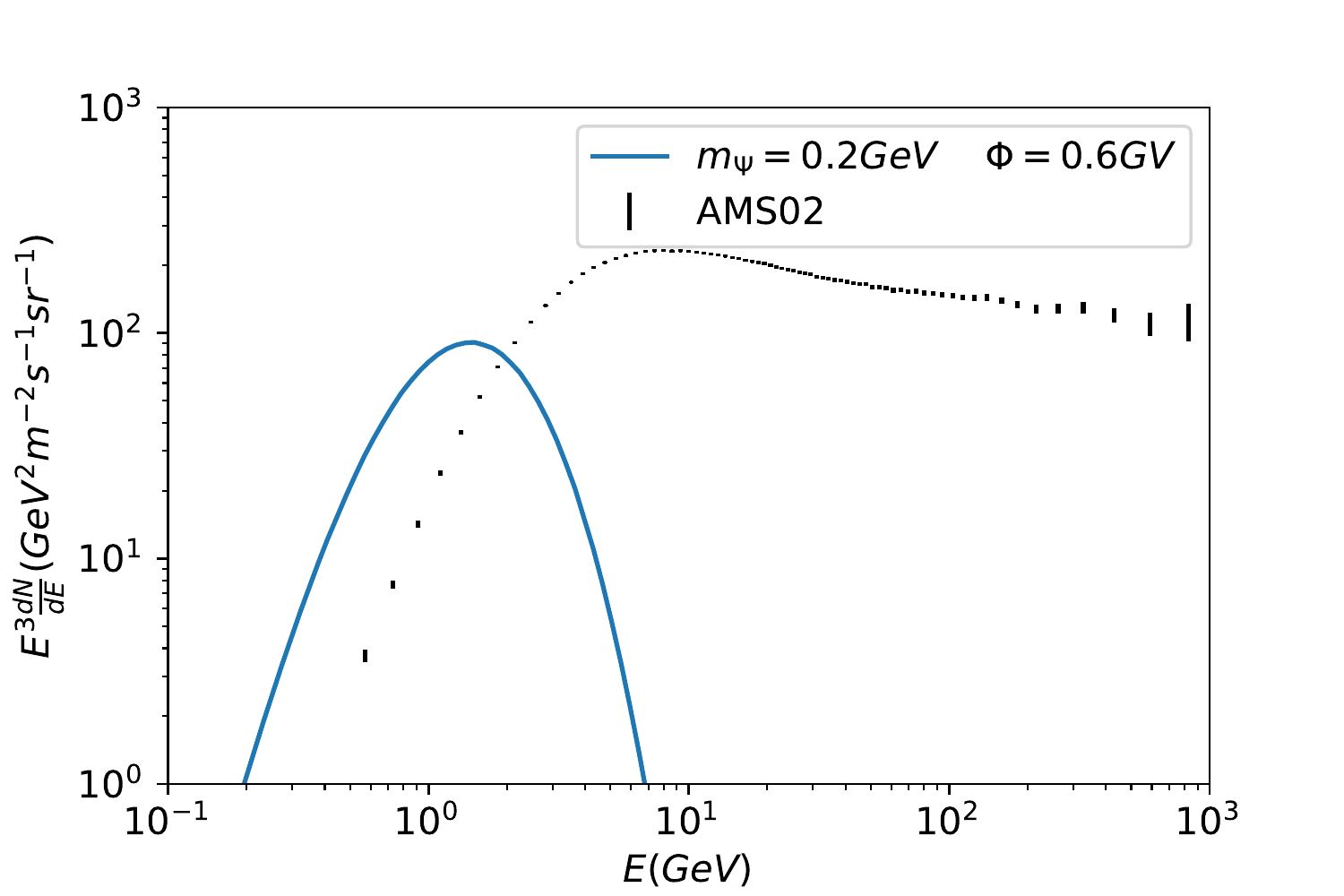}
\caption{Propagated $e^-+e^+$ spectrum from $\Psi\Psi \rightarrow \mu \bar{\mu}$ process, for $m_{\Psi}=0.2$~GeV and $(\sigma v)_{\mu^{-}\mu^{+}}=3\times 10^{-26}$~cm$^{3}$~s$^{-1}$. We assume a solar modulation potential of 0.6~GV.}
\label{propagation}
\end{figure}

Fig.~\ref{propagation} shows the $e^{-}+e^{+}$ spectrum after the propagation, for $m_{\Psi}=0.2$~GeV and $(\sigma v)_{\mu^{-}\mu^{+}}=3\times 10^{-26}$~cm$^{3}$~s$^{-1}$. The AMS-02 measurements are also shown for comparison \cite{Aguilar_2014}. We can see that the reacceleration effect accelerate electrons and positrons to higher energies than $m_{\Psi}$.

\subsection{Background}
Since we focus on the spectral features which are distinct from the ``smooth'' background, it is reasonable to fit the majority of the observational spectra by the background \cite{Bergstrom_2013,Ibarra_2014}. We adopt the background model in \cite{Zu_2017}, which includes three components, the primary $e^{-}$, secondary $e^{-}e^{+}$, and an extra source term of $e^{-}e^{+}$, i.e.,
\begin{eqnarray}
\phi_{e^{-}}&=&C_{e^{-}}E^{-\gamma_1^{e^-}}[1+(E/E_{br}^{e^-})^{\gamma_2^{e^-}}]^{-1}exp(-E/E_{c}^{e^-}),\\
\phi_{e^{+}}&=&C_{e^{+}}E^{-\gamma_1^{e^+}}[1+(E/E_{br}^{e^+})^{\gamma_2^{e^+}}]^{-1},\\
\phi_s&=&C_sE^{-\gamma^s}exp(-E/E_c^s).
\end{eqnarray}
The total background energy spectrum of $e^{-}+e^{+}$ is
\begin{equation}
\phi_{{\rm bkg},e^{\pm}}=\phi_{e^-}+1.6\phi_e^{+}+2\phi_s,
\end{equation}
where the factor 1.6 is due to the asymmetry of the $e^+$ and $e^-$ production in $pp$ collisions \cite{Kamae_2006}.
The best-fit parameters we adopt can be found in Table. I of \cite{Zu_2017}. When we add the DM contribution in the model, we enable the backgrounds to vary to some degree by multiplying adjustment factors $\alpha_iE^{\beta_i}$, with i={$e^{-},e^{+},s$}, on $\phi_e^-,\phi_e^+$, and $\phi_s$, respectively, to optimize the fitting results \cite{Cirelli_2009}.

\section{Results}
\begin{figure}[htbp]
\includegraphics[width=1\columnwidth]{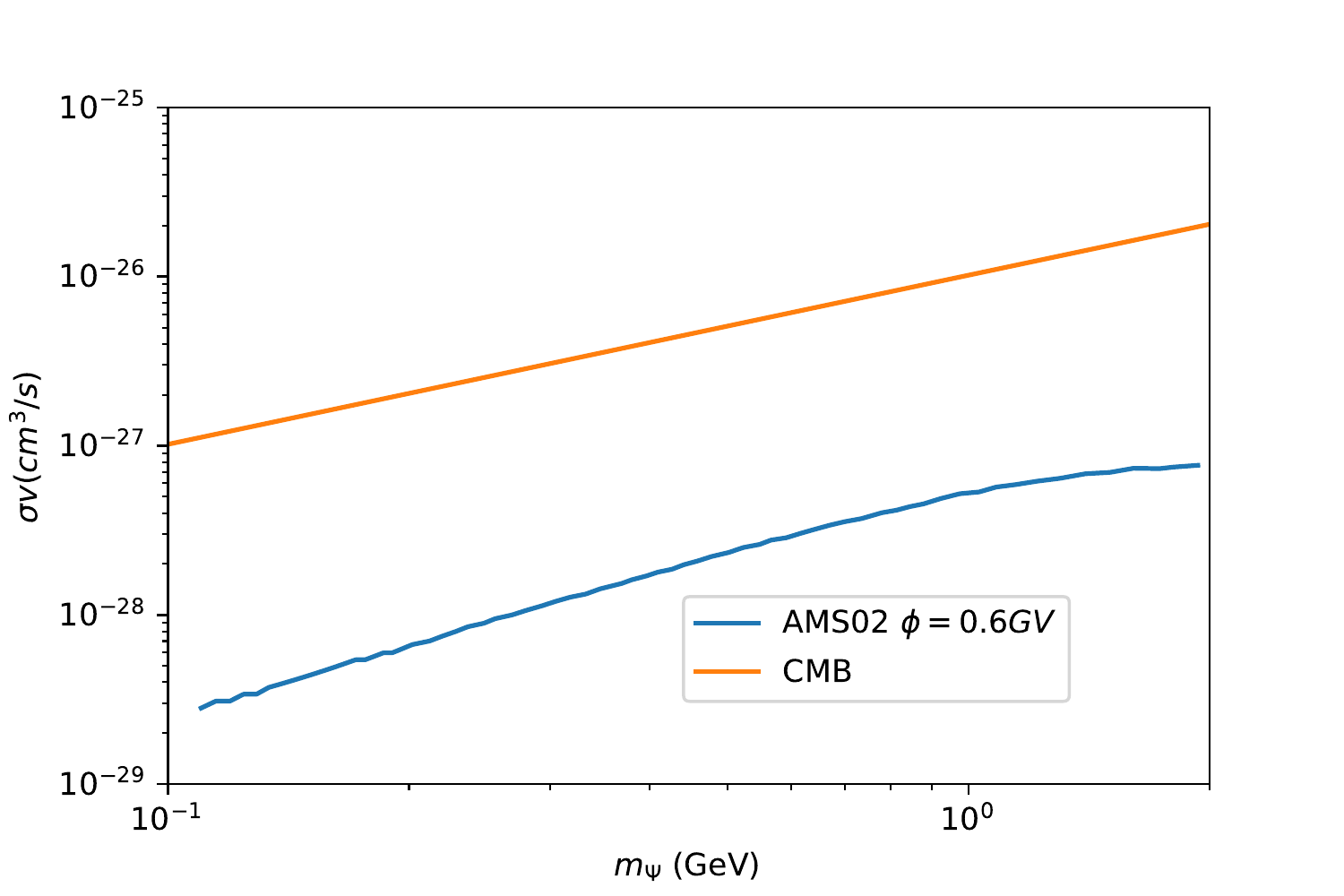}
\caption{The 2$\sigma$ upper limits on the DM annihilation cross section $(\sigma v)_{\mu^{-}\mu^{+}}$ though the process $\Psi \Psi \rightarrow \mu^{-}\mu^{+}$ as a function of DM mass (from 110 MeV to 2 GeV). The conservative CMB limits \cite{Asai_2021} are also shown in orange.}
\label{AMS_uplim}
\end{figure}

We use a maximum likelihood fitting to constrain the DM component. The data used include the AMS-02 positron fraction \cite{Aguilar_2013} and the total electron plus positron flux \cite{Aguilar_2014}. Assuming the DM annihilates to $\mu^{-}\mu^{+}$ in the Milky Way halo, we calculate the $\chi_0^2~(\chi^2)$ without (with) the DM contribution. We set the 2$\sigma$ upper limits on the DM annihilation cross section through setting $\Delta \chi^2=\chi^2-\chi_0^2>2.71$. The results are shown in Fig.~\ref{AMS_uplim}.
The limits on the $(\sigma v)_{\mu^{-}\mu^{+}}$ change from $3 \times 10^{-29}$~cm$^3$~s$^{-1}$ to $7 \times 10^{-28}$~cm$^3$~s$^{-1}$, which are more stringent than the conservative CMB limits \cite{Asai_2021}.

%It needs to notice that our result is actually model independent, if only the DM annihilates through the s-channel to the final state $\mu^{-}\mu^{+}$ only.

\begin{figure}[htbp]
\includegraphics[width=1\columnwidth]{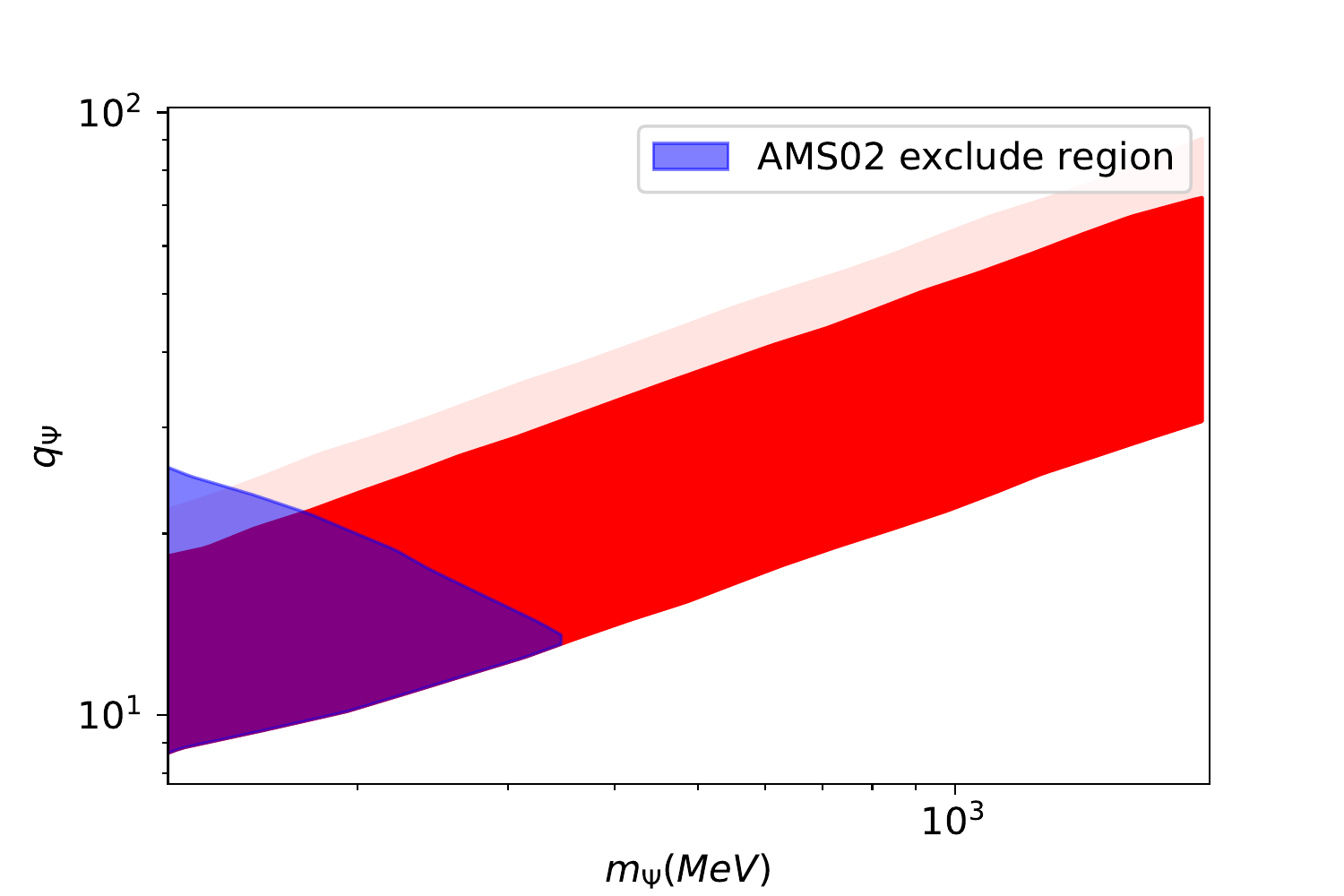}
\caption{The parameter space in the plane of $m_{\Psi}$ and $q_{\Psi}$. The red (light red) region is favored by the $g-2$ at $1\sigma$ ($2\sigma$) confidence level together with the DM relic abundance. The shaded blue region is ruled out by the AMS-02 electron and positron spectra.}
\label{m_q_map}
\end{figure}

When we considering the $L_{\mu}-L_{\tau}$ model, we scan the parameters $m_{X}$ and $g_X$ for any given $m_{\Psi}$ and $q_{\Psi}$ to make sure these parameters are consistent with the current experimental limits. We calculate the $(\sigma v)_{XX}$ for each point to test whether the parameters can explain the DM abundance. The results are shown in Fig. \ref{m_q_map}.

We find that the parameters with 110 MeV$<m_{\Psi}<$400 MeV that can explain simultaneously the $g-2$ anomaly and the DM abundance are partly excluded by the AMS-02 data. The limits from AMS-02 are more stringent in the low mass range. For the $g-2$ anomaly and DM abundance favored regions (red regions), $q_{\Psi}$ is smaller for lighter DM, and thus the $q_{\Psi}^{-2}$ suppression factor of $(\sigma v)_{\mu^{-}\mu^{+}}$ to $(\sigma v)_{XX}$ is bigger. For the same value $(\sigma v)_{XX}$ to explain the relic density, $(\sigma v)_{\mu^{-}\mu^{+}}$ is larger, and hence the AMS-02 constraint is more stringent.

\section{Conclusion}

The $U(1)_{L_{\mu}-L_{\tau}}$ with an MeV scale gauge boson $X$ could explain the $g-2$ anomaly well and avoid the limits from other experiments. A large charged DM that is heavier than the muon and $X$ particle could explain the DM abundance and also escape the constraints from the CMB. In this work, we use the AMS-02 electron and positron data to constrain this model. By means of the reacceleration effect of electrons and positrons in the Milky Way, low-energy electrons and positrons can be accelerated to the AMS-02 energy range, and can thus be strongly constrained. The limits for $(\sigma v)_{\mu^{-}\mu^{+}}$ could be down to $\sim 10^{-29}$~cm$^3$~s$^{-1}$. Part of the parameter region to account for the $g-2$ anomaly can be excluded.

{\bf Acknowledgments} This work is supported by the National Key Research
and Development Program of China (Grant No. 2016YFA0400200), the National
Natural Science Foundation of China (Grants  No. U1738210, No. U1738136, and No. U1738206), Chinese Academy of Sciences, and the Program for Innovative Talents and Entrepreneur in Jiangsu.

\end{document}